\documentclass[12pt]{article} 
\usepackage{graphicx}
\usepackage{color,amsmath,graphicx,latexsym,amssymb} \usepackage{epsfig}
\usepackage{cite} 
\definecolor{li}{rgb}{0,0,1}
\newcommand{\be}{\begin{equation}} \newcommand{\bea}{\begin{eqnarray}}
\newcommand{\bdm}{\begin{displaymath}} \newcommand{\edm}{\end{displaymath}}
\newcommand{\eea}{\end{eqnarray}} \newcommand{\ee}{\end{equation}}
\newcommand{\la}{\langle} \newcommand{\ra}{\rangle}
\newcommand{\Neel}{N\'{e}el } \newcommand{\im}{{\rm i}}
  
\newcommand{\mycomment}[1]{}

\begin{document}
\title{Spiral correlations in frustrated one-dimensional spin-$1/2$ Heisenberg
$J_1$-$J_2$-$J_3$ ferromagnets}

\author
{R.\ Zinke$^1$, J. Richter$^{1}$, and S.-L.\ Drechsler$^{2}$  \\
\small{$^{1}$Institut f\"ur Theoretische Physik, Otto-von-Guericke-Universit\"at
Magdeburg,}\\\small {P.O.\ Box 4120, D-39016 Magdeburg, Germany }\\
\small{$^{2}$ Leibniz-Institut f\"{u}r Festk\"{o}rper- und
Werkstoffforschung (IFW) Dresden,}\\ 
\small{P.O.\ Box 270116, D-01171 Dresden,
Germany}
}
\maketitle
\begin{abstract}
We use the coupled cluster method for infinite chains complemented by exact diagonalization of
finite periodic
chains to discuss the influence of a third-neighbor exchange $J_3$ 
on the ground state of the spin-$\frac{1}{2}$ Heisenberg
chain with ferromagnetic nearest-neighbor interaction $J_1$  and
frustrating antiferromagnetic next-nearest-neighbor interaction $J_2$.
A third-neighbor exchange $J_3$ might be relevant to describe the
magnetic properties of the quasi-one-dimensional edge-shared cuprates, such
as LiVCuO$_4$ or LiCu$_2$O$_2$.
In particular, we calculate the 
critical point  $J_2^c$  as a function of $J_3$,
where the ferromagnetic ground state gives way for a ground state with
incommensurate spiral correlations.
For antiferromagnetic $J_3$ the ferro-spiral transition is always continuous and the
critical values  $J_2^c$ of the classical and the quantum model coincide.
On the other hand, for ferromagnetic $J_3 \lesssim -(0.01 \ldots 0.02)|J_1|$ the
critical value  $J_2^c$ of the quantum model is smaller than that of the
classical model. Moreover, the transition becomes discontinuous, i.e.
the model exhibits a quantum tricritical point.
We also calculate the height of the jump of the spiral pitch angle at the
discontinuous ferro-spiral transition.
\end{abstract}
75.45.+j Macroscopic quantum phenomena in magnetic systems\\
75.10.-b General theory and models of magnetic ordering \\

\section{Introduction}\label{intro}
The recent observation
of spiral (helical) magnetic ground states  in several chain cuprates, such as 
LiVCuO$_4$, LiCu$_2$O$_2$,
NaCu$_2$O$_2$, Li$_2$ZrCuO$_4$, and Li$_2$CuO$_2$
\cite{gibson,matsuda,gippius,ender,drechs1,
drechs3,drechs2,drechs4,park,malek,tarui},
which were identified   as quasi-one-dimensional (1D) frustrated
spin-$1/2$ magnets
with ferromagnetic (FM) nearest-neighbor (NN) in-chain $J_1$ and
antiferromagnetic (AFM)
next-nearest-neighbor (NNN)
in-chain interactions
$J_2$ has stimulated intensive investigations of frustrated 1D
Heisenberg ferromagnets, see, e.g.,
Refs.\cite{rastelli,bursill,honecker,krivnov07,krivnov08,hiki,haertel08,zinke09,lauchli09,sirker2010}.
The 1D $J_1$-$J_2$ model considered in the most theoretical papers may serve
only as the minimal model to describe the magnetic properties of these materials. 
Several extensions, such as exchange  anisotropy \cite{krivnov08,sirker2010} or interchain
coupling\cite{zinke09,ueda09,zhito10,nishimoto10a,nishimoto10b} might be relevant to explain experiments. In
addition to the NN and NNN exchange integrals, $J_1$ and $J_2$, also an 
exchange coupling to 3rd neighbors, i.e. $J_3$, or even couplings to farther distant
neighbors 
could play a role in real materials.  
One mechanism to induce such exchange interactions is strong spin-phonon interaction for
frustrated chains within the antiadiabatic limit \cite{Weisse99}.
Even if these additional couplings are
small, their influence on the spiral ground state (GS) correlations might be
noticeable. In particular, there is a significant influence of $J_3$ on the critical
frustration $J_2^c$ at which the FM GS gives way for the spiral GS, see
below.
Except the possible relevance of the $J_1$-$J_2$-$J_3$ model for real materials the consideration of  
such a model is interesting in its own right as a basic model to study
frustration effects in 1D quantum spin systems.
Moreover, very recently it has been argued that the 
magnetic properties of the kagome-like mineral volborthite
Cu$_3$V$_2$O$_7$(OH)$_2\cdot$2H$_2$O can be described by an effective chain
model with farther distant frustrating exchange couplings\cite{janson}.  
The  corresponding general Heisenberg  Hamiltonian $H$ with NN exchange
$J_1$, NNN exchange $J_2$  
and farther in-chain exchange interactions $J_n$ reads 
\begin{equation}
H=\sum_{n} J_1{\bf s}_n{\bf s}_{n+1}+
J_2{\bf s}_n{\bf s}_{n+2}+
J_3{\bf s}_n{\bf s}_{n+3}+... \quad .
\end{equation}
Motivated by the experiments on the edge-sharing chain cuprates we focus on
the spin-$1/2$ $J_1$-$J_2$-$J_3$ model with FM $J_1$ and frustrating AFM
$J_2 \ge 0$. 
To the best of our knowledge this model has been
investigated so far only in an early paper of Pimpinelli et
al.\cite{rastelli} using spin-wave theory.
   
Here we use the coupled cluster method (CCM) for infinite chains 
complemented by exact diagonalization (ED) of
finite chains (periodic boundary conditions imposed) to investigate spiral GS
correlations. Both methods have been successfully applied to study the spiral
ordering of the $J_1$-$J_2$
model\cite{bursill,aligia00,zinke09}.   
In Refs.\cite{bursill,zinke09} it was demonstrated that the CCM results are 
in good agreement with the DMRG
data. However, in order to take into account the $J_3$ bonds properly, we go
beyond the so-called SUB2-3 approximation used in Refs.
\cite{bursill,zinke09} and consider an improved approximation,
namely the LSUB$4$ approximation,  see below.   
 
The paper is organized as follows. In Sec.~\ref{clas}  
discuss  briefly the classical GS.
In Sec.~\ref{ccm} we provide a brief illustration  of the CCM and describe
its application on the considered model. 
In Sec.~\ref{results} our results for the FM-spiral phase
transition and for the pitch angle in the spiral phase  are presented and
discussed.
In Sec.~\ref{summary} we summarize our findings.

\section{The classical model}\label{clas}
First
we discuss  the GS of the classical model (spin quantum number
$s\to \infty$). 
For the usual $J_1$-$J_2$ model, i.e. the model  with $J_n=0$, $n\geq 3$, studied in many papers  
the critical
frustration $J_2$ is $J_2^c=|J_1|/4$. For $J_2 \ge J_2^c$
there is a spiral GS with a canting angle
between NN (pitch angle) $\gamma$ given by $\cos \gamma =
|J_1|/(4J_2)$.  
This helix interpolates between a
FM chain at $0\leq J_2 \leq J_2^c$ and two decoupled AFM chains
at $J_2/|J_1| =\infty$. Noteworthy, $J_2^c$ is unaffected by quantum
effects, see, e.g., Refs.~\cite{rastelli,bursill,krivnov07,zinke09}. 

If  farther AFM couplings  $J_n$ ($n \geq 3$) are relevant the destabilzation of the
FM GS sets in
for smaller values of $J_2$. 
Extending the classical model including arbitrary $J_n \ne 0$ ($n\geq 3$) one
finds 
for the critical NN exchange 
\begin{equation} \label{eq2}
J_2^c =\frac{1}{4}\left (|J_1|-\sum_{n=3}^\infty n^2 J_n\right). 
\end{equation}
This expression has been derived assuming a {\it continuous}
transition between the FM and the spiral GS's.
It holds for arbitrary AFM
long-range
couplings $J_n$ with $n \geq 3$. 
In case that some of the exchange couplings
are FM, i.e. $J_n<0$ for certain $n \ge 3$,  
the Eq.~(\ref{eq2}) holds only  if the AFM couplings dominate. 
%
Assuming
$J_2> 0$ we find as the criterion for the validity of Eq.~(\ref{eq2})  
\begin{equation} \label{eq_a}
J_2 \geq  -\frac{1}{12}\sum_{n=3}^{\infty}n^2\left( n^2-1\right)J_n,
\end{equation}
or equivalently 
\begin{equation} \label{eq_b}
|J_1|  \geq 
-\frac{1}{3}\sum_{n=3}^{\infty}n^2\left( n^2-4 \right)J_n.
\end{equation}
If this condition is violated, in crossing the FM-spiral phase boundary, 
the spiral GS ''jumps'' from a {\it finite}
 pitch angle $\gamma_T \neq 0$ 
to $\gamma=0$
in the FM GS. 
For the simplest case $J_3 \neq 0$ and $J_n=0$ ($n > 3$) the classical model
was considered by  Pimpinelli et
al.\cite{rastelli}. They found  for the critical frustration $J_2$ in case of
continuous transition $J^c_2=(|J_1|-9J_3)/4 $ in accordance with the general
expression (\ref{eq2}). 
The classical pitch angle in the
spiral phase is given by
$\cos \gamma =
\left (-J_2 +\sqrt{J_2^2+3J_3(3J_3+|J_1|)}\right )/6J_3$.
According to the general Eqs.~(\ref{eq_a}) and (\ref{eq_b})
for the $J_1$-$J_2$-$J_3$ model the classical FM-spiral transition
is  discontinuous at  $J_3 < -\frac{|J_1|}{15}$ 
or equivalently at $J_3 < -\frac{J_2}{6}$.

For $J_3 < 0$ and $J_n=0$, $n > 3$,
the height of the jump of the pitch angle $\gamma_T$ at the transition is given
by\cite{rastelli}
\begin{equation}\label{eq_Nr_1}
\cos \gamma_T= -J_1\left(2J_3+\sqrt{4J_3^2+4J_3J_1}\right)^{-1} -\frac{3}{2} 
\end{equation}
or equivalently 
\begin{equation} \label{eq_Nr2}
\cos \gamma_T= \frac{-J_2}{4J_3}-\frac{1}{2}    . 
\end{equation}
Considering other simplified classical models with a single FM long-range
coupling $J_{n_0}$, i.e. a model with $J_1<0$, $J_2> 0$,
$J_{n_0} < 0$, $J_n=0$ ($n > 2$ and $n\ne n_0$),
the classical discontinuous FM-spiral transitions occurs according to 
Eqs.~(\ref{eq_a}) and (\ref{eq_b}) at
$-J_{n_0} > 12J_2/\left(n_0^2 \left(n_0^2-1\right)\right)$ (or
equivalently
at $-J_{n_0} > 3|J_1|/(n_0^2(n_0^2-4))$. 
In other words, even a tiny but fairly long range ferromagnetic coupling may introduce a 
discontinuous behavior at the critical point.

\section{The Coupled Cluster Method (CCM)}\label{ccm}
For the sake of brevity, we will outline only some important
features of the CCM which are relevant for
the model under consideration.
The interested reader can find more details
concerning the application of the CCM on the frustrated Heisenberg magnets
with non-collinear GS's in
Refs.~\cite{bursill,krueger00,krueger01,farnell01,ivanov02,rachid05,rachid06,darradi08,zinke09,bishop09,ccm_m_h_09}. 
For more general aspects of
the methodology of the CCM, see, e.g.,
Refs.~\cite{zeng98,bishop00,farnell04}.

First we mention that the CCM approach 
yields results in 
the thermodynamic limit $N\to\infty$.  The starting point for a CCM
calculation is the choice of a normalized reference (or model) state
$|\Phi\rangle$. Related to this reference state we then define a set of 
mutually commuting multispin
creation operators $C_I^+$, which are 
themselves defined over a complete set of
many-body configurations $I$.  
For the considered frustrated spin system we choose a spiral
reference state with spiral spin orientations along the chains (i.e., pictorially, $|\Phi\rangle
=|\uparrow\nearrow\rightarrow\searrow\downarrow\swarrow\cdots\rangle$)
characterized by a pitch angle $\gamma$, i.e.
$|\Phi\rangle=|\Phi(\gamma)\rangle$.
Such states include the FM state ($\gamma=0$) as well as the \Neel state
($\gamma=\pi$).
Next, we perform a rotation of the local axis of
the spins such that all spins in the reference state align along the
negative $z$ axis.  
This rotation by an appropriate local angle $\delta_{n}$ of the 
spin on lattice site $i$
is equivalent to the spin-operator transformation
\begin{equation}
\label{eq3} 
s_{n}^x = \cos\delta_{n} 
{\hat s}_{n}^x+\sin\delta_{n} {\hat s}_{n}^z\; ;\; 
 s_{n}^y = {\hat s}_{n}^y   
\; ;\;s_{n}^z = 
-\sin\delta_{n} {\hat s}_{n}^x+\cos\delta_{n} {\hat s}_{n}^z ,
\end{equation}
where ${\hat s}_{n}^x$, ${\hat s}_{n}^y$, ${\hat s}_{n}^z$ are the spin
operators in the rotated coordinate frame.
The local rotation angle $\delta_{n}$ is related to the 
pitch angle $\gamma$ of the spiral reference state by $\delta_{n}=n\gamma $.
In this new set of local spin coordinates 
the reference state and the corresponding multispin
creation operators $C_I^+$ are  given by
\begin{equation}
\label{set1} |{\hat \Phi}\ra = |\downarrow\downarrow\downarrow\cdots\rangle ; \mbox{ }
C_I^+ 
= {\hat s}_{n}^+ \, , \, {\hat s}_{n}^+{\hat s}_{m}^+ \, , \, {\hat s}_{n}^+{\hat s}_{m}^+{\hat
s}_{k}^+ \, , \, \ldots \; ,
\end{equation}
where the indices $n,m,k,\ldots$ denote arbitrary lattice sites.
In the rotated coordinate frame the Hamiltonian
becomes dependent on the pitch angle $\gamma$. It reads
\begin{eqnarray}\label{hamiltonian_trafo}
H&=& \sum_{m=1}^3  \frac{J_m}{4}\sum_{n} [\cos(m\gamma)+1]({\hat s}_{n}^+{\hat
s}_{n+m}^-+{\hat s}_{n}^-{\hat s}_{n+m}^+)\nonumber \\
& + &
[\cos(m\gamma)-1]({\hat s}_{n}^+{\hat s}_{n+m}^++{\hat s}_{n}^-{\hat s}_{n+m}^-) + 2\sin(m\gamma)[{\hat s}_{n}^+{\hat
s}_{n+m}^z \nonumber \\  
&-&   {\hat s}_{n}^z {\hat s}_{n+m}^+  +{\hat s}_{n}^-{\hat
s}_{n+m}^z-{\hat s}_{n}^z{\hat s}_{n+m}^-]+4\cos(m\gamma){\hat s}_{n}^z{\hat
s}_{n+m}^z,
\end{eqnarray}
where ${\hat s}_{i,n}^{\pm}\equiv {\hat s}_{i,n}^x\pm \im {\hat s}_{i,n}^y$. 

With the set $\{|\Phi\rangle, C_I^+\}$ the CCM parametrization of 
the exact ket 
GS eigenvector
$|\Psi\ra$ 
of the many-body system 
is given  by
\begin{equation}
\label{eq5} 
|\Psi\ra=e^S|\Phi\ra \; , \mbox{ } S=\sum_{I\neq 0}a_IC_I^+ .
\end{equation}
The CCM correlation operator $S$ contains the correlation coefficients
$a_I$, 
which  can be determined by the so-called set of the CCM ket-state 
equations
\begin{equation}
\label{eq6}
\langle\Phi|C_I^-e^{-S}He^S|\Phi\rangle = 0 \;\; ; \; \forall I\neq 0 , 
\end{equation}
where $C_I^-=\left (C_I^+ \right )^{\dagger}$.
Each ket-state 
equation belongs to a specific creation operator
$C_I^+=s_n^+,\,\,s_n^+s_{m}^+,\,\, s_n^+s_{m}^+s_{k}^+,\cdots$,
i.e. it corresponds to a specific  set (configuration) of lattice sites
$n,m,k,\dots\;$.
By using the Schr\"odinger equation, $H|\Psi\ra=E|\Psi\ra$,
we can write the GS energy as $E=\la\Phi|e^{-S}He^S|\Phi\ra =E(\gamma)$, which depends
(in a certain CCM approximation, see below)
on the pitch angle $\gamma$.  
\\In the quantum model the pitch angle may be different from the corresponding classical value $\gamma_{\rm cl}$. 
Therefore, we do not choose the classical  result for the pitch angle in the quantum model, rather, we consider $\gamma$ as a free
parameter in the CCM calculation, which has to be determined by minimization of the
CCM GS energy $E(\gamma)$, i.e.
$dE/d\gamma|_{\gamma=\gamma_{qu}}=0$.
\\For the many-body quantum system under consideration
it is necessary to use approximation 
schemes in order to truncate the expansions of $S$ 
in  Eq.~(\ref{eq5}) in a practical calculation.
In Refs.~\cite {bursill} and \cite{zinke09} it has been demonstrated, 
that  for
the $J_1$-$J_2$ model the so-called SUB2-3 approximation
leads to results of comparable accuracy to those obtained using the DMRG
method \cite{white96}.
In this approximation all configurations are included which span a
range of no more than
$3$ contiguous sites and contain only  $2$ or fewer spins.
Taking into account the $J_3$ bond we have to extent this approximation in
order to take into account configurations including a
range of $4$ contiguous sites. The corresponding approximation is the
so-called  
LSUB4 approximation, see
e.g., Refs.~\cite{bishop00,farnell04,darradi08}. Within this approximation 
multispin creation operators
of one, two, three or four spins distributed on clusters of
four contiguous lattice sites are included.

In addition, for the determination of the quantum tricitical point 
we have also used higher LSUB$n$  approximations, see
Sec.~\ref{results}. However, 
the numerical complexity increases tremendously, since (i) the number of
ket-state equations (\ref{eq6}) increases exponentially with $n$, 
(ii) there are two free parameters $J_2$, $J_3$ 
which have to be varied by very small increments to find the transition
points, and 
(iii) the determination of the quantum pitch angle $\gamma_{qu}$ 
itself requires the iterative minimization of the ground state energy for each set of $J_2$, $J_3$.
Hence, except for the determination of the quantum tricitical point, we have
restricted our CCM calculations to the CCM-LSUB4 approximation.
\begin{figure}
\scalebox{1.2}{\includegraphics{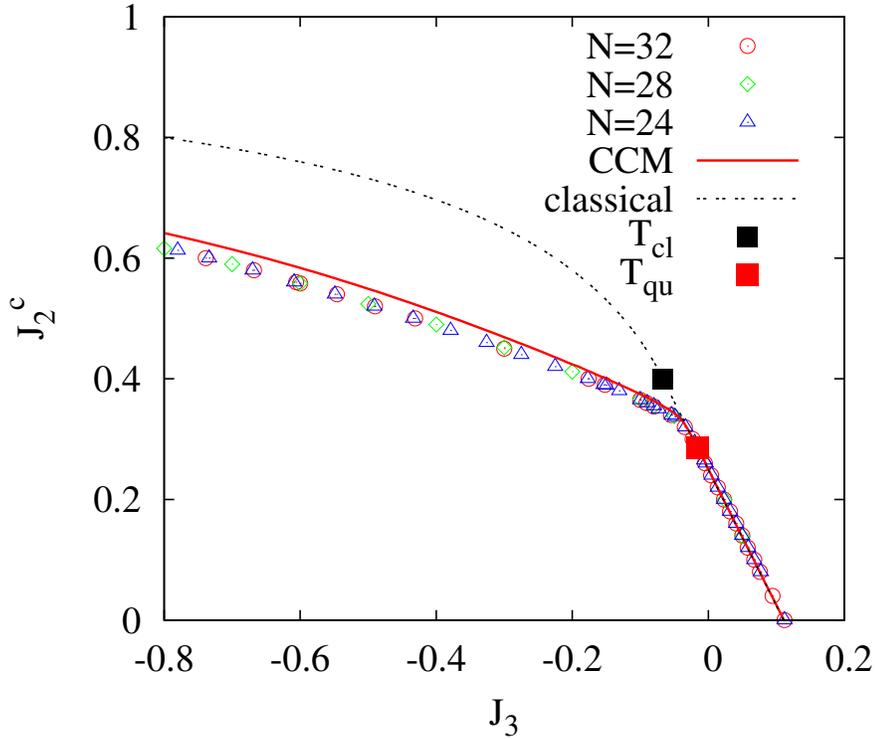}} 
\caption{Phase diagram of the $J_1$-$J_2$-$J_3$ Heisenberg
chain determined by CCM for $N\to \infty$ and ED for $N=24, 28, 32$.
Below the transition line the GS is the fully polarized
FM state. Above the line the GS exhibits spiral correlations.
The transition can be continuous (larger $J_3$) or discontinuous (smaller
$J_3$), see text.
Note that the classical transition line (black dashed) corresponds to the
result of Pimpinelli et
al.\cite{rastelli}. 
}\label{trans}
\end{figure}

\section{Results for the spin-$1/2$ quantum model}\label{results}

In what follows we set $J_1=-1$ if not stated otherwise explicitly.
The phase diagram of the $J_1$-$J_2$-$J_3$ chain with FM $J_1=-1$
obtained by the CCM and by the ED
is shown in Fig.~\ref{trans}. For the classical as well as for the quantum
model the transition from the FM to the spiral GS can be 
second ($J_3 > J_3^{T}$)
or first order ($J_3 < J_3^{T}$), i.e.,
the pitch angle $\gamma$ does (first order) or does not (second order) jump from $\gamma=0$ to a finite
value $\gamma=\gamma_T
>0$ at the transition. 
The tricritical point $T=(J_2^{T},J_3^{T})$, i.e. 
that point at the transition line where the second-order transition  
goes over in a first-order transition is
$T_{cl}=(J_2^{T_{cl}},J_3^{T_{cl}})=(\frac{2}{5},-\frac{1}{15})$ for the classical model
\cite{rastelli}, see the black square in Fig.~\ref{trans}.
Due to quantum fluctuations this point is shifted to 
smaller
values of $|J_3|$.
The LSUB4-CCM estimate for the  quantum tricritical point is   
$T_{qu}^{LSUB4}=(J_2^{T_{qu}^{LSUB4}},J_3^{T_{qu}^{LSUB4}}) = (0.283,-0.013)$, see the red square in
Fig.~\ref{trans}.
This result is quite close to the spin-wave
result \cite{rastelli}
$T^{sw}_{qu}=(J_2^{T_{qu}^{sw}},J_3^{T_{qu}^{sw}})=(0.25,0)$. 
We have determined the quantum tricritical point also 
using  higher CCM-LSUB$n$ approximations, namely LSUB6, LSUB8 and LSUB10.
The corresponding results are
$(J_2^{T_{qu}^{LSUB6}},J_3^{T_{qu}^{LSUB6}}) =
(0.274,-0.011)$, 
$(J_2^{T_{qu}^{LSUB8}},J_3^{T_{qu}^{LSUB8}})\\ = (0.268,-0.009)$, and
$(J_2^{T_{qu}^{LSUB10}},J_3^{T_{qu}^{LSUB10}}) = (0.266,-0.007)$. Obviously,
these values are quite close to each other, and there is a slight shift
towards $(J_2^{T_{qu}^{sw}},J_3^{T_{qu}^{sw}})$ of Ref.~\cite{rastelli}
when increasing the order of CCM approximation.

The ED data for the transition line are in very good agreement with the CCM
data. There is only a slight deviation for stronger FM $J_3$ bonds, i.e. for
$J_3 \lesssim -0.2$.
Note, however, that in a tiny interval
$-0.1 \lesssim J_3 \lesssim -0.05$ for finite chains the transition from the fully polarized
FM state with total spin $S=N/2$ is not directly to a singlet state
with $S=0$, rather there are
some intermediate states  with $N/2 > S > 0$.
To give an example, for $J_3=-0.08$ and $N=32$ there is a sequence of
transitions  from the FM state to a singlet GS via partially
polarized states in the interval $0.354 < J_2 < 0.372$.
The estimation of the 
the tricritical point by ED is
$T^{ED}_{qu}=(J_2^{T_{qu}^{ED}},J_3^{T_{qu}^{ED}}) \approx (0.28,-0.01)$,
cf. Fig.~\ref{fig4}.  
Note that for $J_3>J_3^{T_{qu}}$ the classical and quantum transition lines
coincide, i.e. the relation  $J^c_2=(1-9J_3)/4 $ is valid also for the
quantum model. Due to the quite large prefactor $-9/4$, already a
         weak 3rd neighbor coupling $J_3$ has a noticeable effect on
the critical $J^c_2$. For $J_3<J_3^{T_{qu}}$ the transition point of the
quantum model is shifted
to smaller values of frustrating $J_2$. 
That is to some extent surprising, since in
most of the previous studies of models exhibiting a transition
between a spiral and a collinear GS in the quantum model the opposite
behavior has been found, i.e., the transition to the spiral GS is
shifted to higher values of frustration, see, e.g.,
Refs.~\cite{white96,aligia00,bursill,krueger00,rachid05,zinke09,bishop09}.

\begin{figure}[htp]
\scalebox{1.0}{\includegraphics{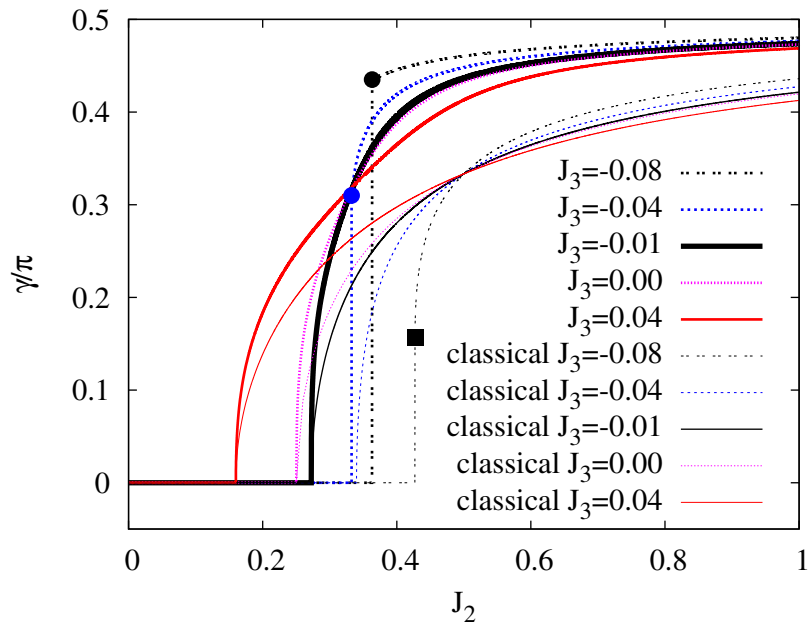}} 
\caption{\label{gamma}The CCM results for the quantum pitch angle
$\gamma_{qu}$ in dependence on $J_2$ for various values of $J_3$. For comparison we also show
the classical pitch angle $\gamma_{cl}$. Note that for $J_3=-0.08$ (quantum and classical model) and for $J_3=-0.04$ 
(quantum model only) the pitch angle jumps from zero to a finite value $\gamma_T$ at the transition point. These jumps are
indicated by filled circles (quantum model) or a filled  square (classical model).}
\end{figure}
To illustrate the quantum tricritical point in more detail we show
in Fig.~\ref{gamma}  the CCM results for the pitch angle versus $J_2$ for various values of
$J_3$ around $J_3^{T_{qu}^{LSUB4}}$. For comparison we show also the classic pitch
angle $\gamma_{cl}$.
It can be clearly seen how the continuous behavior of the
pitch angle $\gamma$ goes over
into a discontinuous one.
Interestingly, at a particular value of $J_2=J_2^*$ the curves cross
each other, i.e. the pitch angle is independent of $J_3$. 
For the classical model the crossing point is at
$J_2^*=1/2$ and the corresponding pitch angle is $\gamma_{cl}=\pi/3$.
For the quantum model the curves do not cross exactly in a  point, rather
they approach each
other very closely at $J_2^*=0.335$. The pitch angle at that point is
$\gamma_{qu}=0.32\pi$. 
Furthermore the quantum pitch angle $\gamma_{qu}$ approaches the limiting
value  $\lim_{J_2\to \infty} \gamma_{qu}$ for much smaller values of $J_2$
than the  classical one.

\begin{figure}[htp]
\scalebox{1.2}{\includegraphics{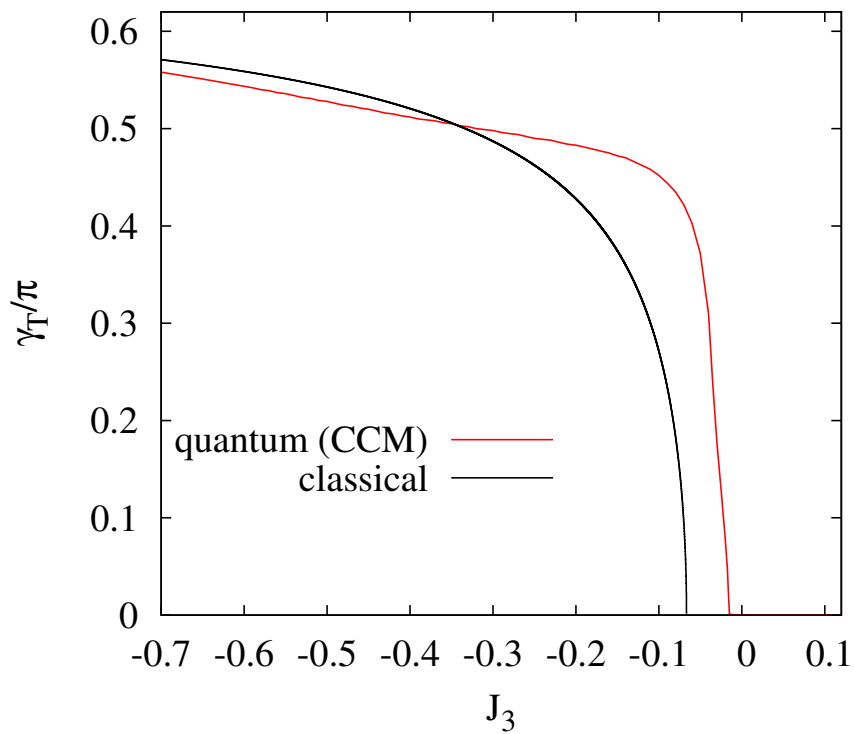}} 
\caption{\label{fig3}The height of jump of the pitch angle $\gamma_T$ at the
transition point in dependence on $J_3$.}
\end{figure}

\begin{figure}[htp]
\scalebox{1.2}{\includegraphics{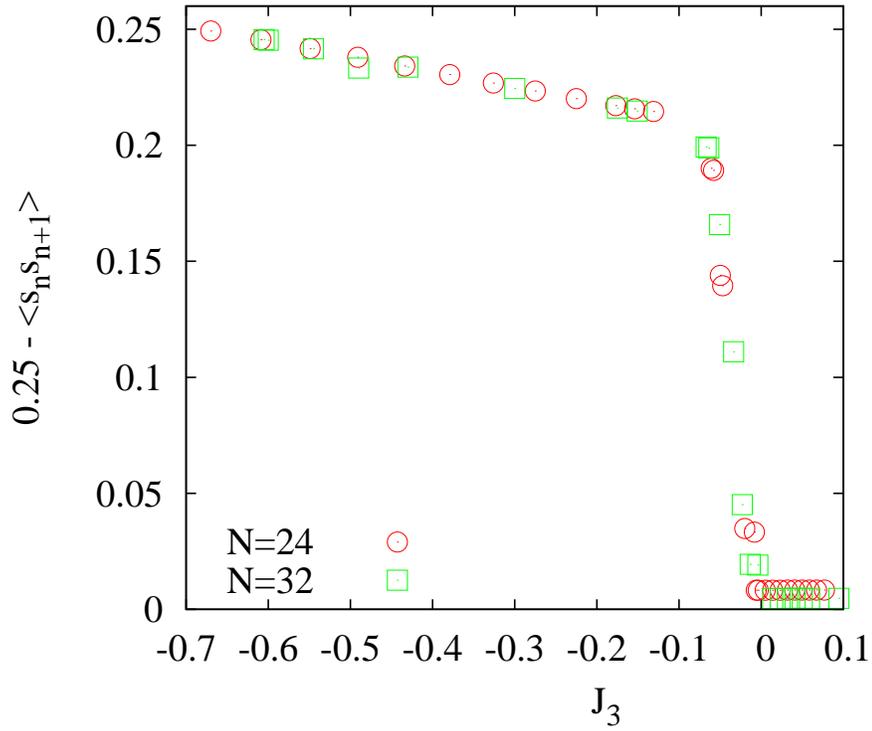}} 
\caption{\label{fig4} The jump of the nearest-neighbor spin-spin correlation
function at the transition from the FM to the singlet GS, 
cf. Fig.~\ref{trans}, for finite periodic chains of length N=24 and N=32.}
\end{figure}

In Fig.~\ref{fig3} we present the height of the jump $\gamma_T$ at the transition
point in dependence on $J_3$.  For the classical model $\gamma_T$ is given by
Eq.(\ref{eq_Nr_1}) for $J_3<-1/15$. 
For the quantum model the $\gamma_T(J_3)$ curve is characterized by two
nearly linear regimes, one regime (near the quantum tricritical point) 
with a steep increase of $\gamma_T$ 
and a second, almost flat one for $J_3
\lesssim -0.2$. 
This scenario is confirmed by the ED results for the NN spin-spin
correlation function shown in Fig.~\ref{fig4} which may serve as a finite-chain
analogue of the
infinite-chain pitch angle.

Finally, let us discuss the pitch  angle which appears in the limits of large  $J_2$ or large
$| J_3 |$. This limiting value of $\gamma $ is the maximal pitch
angle and it is monotonously
approached from below increasing the corresponding bond $J_2$ or $|J_3|$ while
fixing the other one.    
For $J_2 \to \infty $ (and finite $|J_3|$) the pitch angle approaches $\gamma
=\pi/2$. In this limit
the system splits into two
decoupled AFM chains with coupling strength $J_2$. 

For $J_3 \to \infty $ (and finite $J_2$) the pitch angle approaches $\gamma=
\pi/3$, i.e. only acute pitch angles appear.
For $J_3 \to -\infty $ (and finite $J_2$) the pitch angle approaches $\gamma=
2\pi/3$, i.e. by contrast to the pure $J_1$-$J_2$ model also
obtuse pitch angles appear. As it is obvious from Fig.~\ref{fig3} pitch angles $\gamma >
\pi/2$ appear already for quite moderate values of $J_3$.  
We mention that the above discussion is not
purely academic, since at least values $J_2 > 1$ 
might be realized also 
in real
materials, e.g.\ in NaCu$_2$O$_2$\cite{drechs3}.


\section{Summary}\label{summary}

Using the coupled-cluster method (CCM) and the Lanczos exact diagonalization
technique (ED) we have studied the influence of a third-neighbor
exchange $J_3$
on the GS of the spin-$\frac{1}{2}$ Heisenberg
chain with FM NN interaction $J_1$  and
frustrating AFM NNN interaction $J_2$. In particular, we have analyzed the 
transition from the FM GS (present for dominating $J_1$) to a
singlet GS with incommensurate short-range spiral correlations.  
The  results obtained by these two different approximations agree well.
Moreover, the finite-size effects inherent in the ED study appeared to be
small. 

We have found, 
that  in case of an AFM coupling $J_3$ the FM-spiral  transition point $J_2^c$ of the quantum model
coincides with that of  the classical model, and it is always continuous.
However, the quantum pitch angle significantly deviates from the classical one. 
For  a FM coupling $J_3$ quantum fluctuations shift 
the FM-spiral  transition point $J_2^c$ to smaller values, and
the transition becomes discontinuous.\\

{\bf Acknowledgment}
We thank the
DFG for financial support (grants 
DR269/3-1  and RI615/16-1).
For the exact diagonalization  J.~Schulenburg's {\it
spinpack} was used.

\end{document}